# Open Access
# all you wanted to know and never dared to ask

**Alfonso Pierantonio**[*], **Mark van den Brand**[†], and **Benoit Combemale**[‡]
[*]Università degli Studi dell'Aquila, Italy
[†]Eindhoven University of Technology, The Netherlands
[‡]University of Toulouse & Inria, France

**ABSTRACT** This editorial presents the various forms of open access, discusses their pros and cons from the perspective of the Journal of Object Technology and its editors in chiefs, and illustrates how JOT implements a platinum open access model. The regular reader will also notice that this editorial features a new template for the journal that will be used from now on.

## 1. Introduction

Over the last years, the advent of Open Access (OA) is inducing profound changes in scientific publishing. The impact of such transformation is comparable to that of the introduction of movable type printing (AD 1455), and the transition to electronic publishing. OA is a mechanism that allows to openly and immediately access without costs for the consumer research results and data. The basic principle is that the research products financially supported by public research programs and agencies must be openly accessible. The objectives include

– enhancing the dissemination on a global scale,
– making research products accessible to people who do not have access to paywall-based distribution systems,
– reducing research duplication,
– supporting interdisciplinary research, knowledge transfer, and making the research process more transparent to the taxpayer,
– increasing the use of scientific contributions in teaching programs, and
– making research results perpetual.

Thus, conventional wisdom suggests that OA can only have a beneficial influence on academic communities and the dissemination of knowledge and resources among researchers. However, a recent survey (Morais and Borrell-Damian 2018) conducted by the European University Association[2] documented a significant lack of awareness among academic people, as illustrated in Fig. 1. Librarians are much informed about the current movement towards more liberal access to knowledge. In this survey, researchers (at any stage of their career) seem to have a less proactive attitude within their institutions and their corresponding community, although individual communities developed an internal debate on the topic, e.g., the European Mathematical Society[3]. In particular, librarians and institutional leaders have the best knowledge of the OA rules. Consequently, it is not easy for researchers to make informed decisions leaving the decision-making process in the hands of those who are informed most and becoming somewhat unable to influence something that is going to change their professional life. Moreover, shifting the reader's cost to the authors can be conducive to conflicts and financial bias because OA removes the barriers to access to research products without eliminating the publication costs, i.e., the so-called Article Processing Charge (APC). The risk is that the publishers will drive this epochal change leaving to the primary actors, i.e., the researchers, with little or no control over it. In particular,

a) publishers might be induced to publish more because for each accepted paper there is an APC to pay that may top several thousands euro depending on the discipline, and
b) specific researchers can be invited to join the authors' list

---
[2] https://www.eua.eu/
[3] https://euro-math-soc.eu/system/files/news/EMS-PED-OA-PlanS.pdf

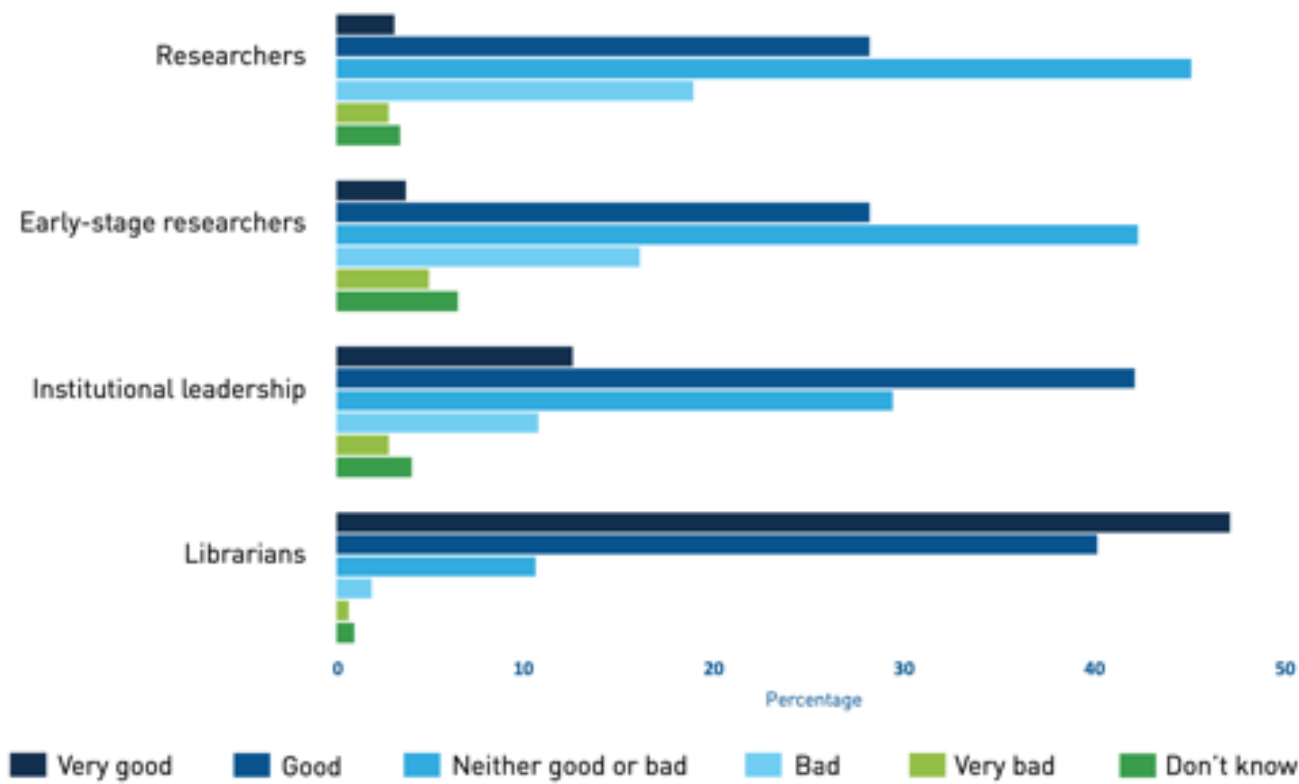

**Figure 1** Awareness of scientific publishers' open access policies among different university populations ([Morais and Borrell-Damian 2018](#))

of a paper only because their institutions can cover the APC by means of *transformative agreements*.

Such threats represent a financial bias that is exacerbated by the economic conditions of the social, industrial, and academic contexts. Furthermore, they might cause a range of conflicts at both local and global levels. Readers and authors within the same department, institution, or country may not overlap, requiring the academic system to reallocate part of the library funds to allow authors to cover the publication costs.

Analyzing the aspects that can limit a researcher's ability to publish and conduct an independent line of research freely is critical as it pertains to the ethical values of individuals at different levels in the organizational hierarchy of academic institutions. Necessarily, it is vital to understand the challenges and the opportunities that OA may give place.

## 2. The Open Access models

In this section, the different OA models are presented.

**Hybrid Open Access.** "Hybrid" open access means that one or more articles in a subscription journal may be open to anyone on the internet even though all the rest of the content is available only to people and institutions with paid subscriptions. This is possible in some journals that offer their authors a choice of paying the APC to make their article freely accessible or leaving it behind the subscription barrier.

**Green Open Access.** "Green" open access occurs when the publisher of a subscription journal allows the author to keep the non-commercial rights to her article so it can be posted in open internet archives. Archives may be institutional repositories or discipline-specific archives maintained by scholarly associations. In some cases, the publisher requires that open access in the archives be delayed for 6 to 12 months.

A variety of platforms are available, including disciplinary repositories (e.g., https://arxiv.org, PubMed Central) and institutional repositories hosted by a university or organization (e.g., HAL). The benefit of green open access for researchers is the avoidance of costs.

**Gold Open Access.** "Gold" open access refers to journals in which all articles and content are open access — available to anyone on the internet without any subscription fees or sign-in. Publishers cover the publishing costs by means of the APC, which is charged to authors to make a work available open access. This fee is usually paid by an author's institution or research funder rather than by the author themselves. An article processing charge does not guarantee that the author retains copyright to the work, or that it will be made available under a Creative Commons license (see below).

**Platinum Open Access.** "Platinum" (a.k.a. Diamond) open access means permanent and free access to published scientific works for readers with no publication fees for the authors – 100% free. All articles are published under the most flexi-ble reuse standard – the CC BY license. Publication costs are funded by non-profit associations, academic institutions or governmental agencies. Platinum open access platforms maximise the potential for exchanging ideas and provides a valuable contribution to those with limited financial means by levelling the playing field and giving everyone an equal chance to publish and read scientific publications of high quality.

**Black Open Access.** Black open access refers to a recent development leading to the emergence of other potentially illegal channels for uploading and accessing research articles without subscriptions, payments, and bureaucracy. The most notable example of such platforms is Sci-Hub. The repository has been populated with more than 50 million journal articles retrieved directly from the publishers' websites (Bohannon 2016). It is unclear whether this has been realized by using passwords that have been freely made available or obtained with phishing.

Interestingly, the accesses to Sci-Hub are not limited to scholars in less developed countries. Sci-Hub had 28 million download requests in between September 2015 through February 2018, from all regions of the world and covering most scientific disciplines (Bohannon 2016). What makes Sci-Hub unique is that it combines OA and indexing in the same platform: not even Google Scholar can do that! In essence, the paywall model jeopardizes usability and immediacy.

## 3. Scientific Publishing Market

The market is defined by the goods that are exchanged by the actors operating in it. In scientific publishing, the good exchanged is a particular one, knowledge! Historically, most production costs were variable costs, depending on the number of copies to be produced and the number of copies to be distributed. Today, the cost is fixed and is related to the processing, production, and distribution platform, but the reproduction costs are nonexistent. What publishers are selling can be summarized as follows:

– contents,
– organization, and
– certification.

The contents are the research products developed by researchers, reviewed by other researchers, and managed by the editor-in-chief and the editorial board of the journal. The organization is mainly referring to the activities, e.g., sending reminders to late reviewers, content collections, and finally the platform, which is probably the most expensive task (besides contractual and administrative activities) afforded by the publishers. Finally, the certification represented by the consolidated reputation over the years of published journals and those involved in their lifecycle. The question is how to clearly identify the value publishers are adding to the scientific production and whether it is useful or needed (Van 2013).

The scientific publishing market is also characterized by its inelasticity, i.e., the products cannot be easily replaced. If an individual is looking for a given paper, there are little or no chances that she will find what needed in another publication: papers are not replaceable, and the publishers have the power to influence the price, directly or indirectly, something that does not occur under perfect competition as demonstrated by increasingly higher costs that do not correspond to the increase of the number of published papers[4].

In this context, with a limited number of global players influencing the market prices, the idea that universities will take the funds they spend on subscriptions, and gradually flip that cash toward transformative agreements, i.e., hybrid deals that pay for both subscriptions and APCs, can be deceptive.

## 4. There is a catch!

Gold OA publications are typically operated on commercial platforms with the APC shifting the burden of payment from readers to authors (or their funders). Therefore, Gold OA introduces perverse financial incentives on both sides of the knowledge publication process: publishers might be tempted to increase the number of accepted publications, and financial barriers might be introduced among scientists including authors from less rich countries, institutions, or communities. Moreover, institutional budgets may need to be adjusted in order to provide funding for the article processing charges required to publish in many open access journals, while retaining the old subscriptions at the same time. This is often mitigated by the so-called transformative agreements that, however, are difficult to be finalized (see https://sparcopen.org/our-work/big-deal-cancellation-tracking/), very often include packages of journals ("big deals") that may not always be of interest, and especially they are often managed at national level.

In this complex scenario, the schedule imposed by Plan-S[5] is also going to have an impact. Starting from January 2021, all scientific publications that result from research funded by public grants must be published in compliant Open Access journals and platforms. Without going too much into detail about Plan-S, scientific communities might suffer a lack of representativeness as their traditional, long-standing, subscription-based scientific journals may not easily conform to Plan-S.

## 5. Mutualizing Costs at JOT

JOT is a Platinum Open Access[6] journal. Readers can freely access the published manuscripts on the journal platform, whereas authors do not have to sustain the processing costs. The journal runs on the voluntary contribution of authors, reviewers, editorial board members, and editors as in any other scientific journal. In Table 1, a non-exhaustive list of the activities covered voluntarily is summarized. It is worth noting that a journal lifecycle is made of many activities ranging from trivial tasks like printing the certificates to be issued in favor or those who helped over time to more complex functions like maintaining and upgrading the submission system. All have something in common, no matter how relevant they are, all require dedication, are time-consuming, and often are preemptive and must be prioritized. The only way this is made possible is to have the firm and steady support of the community partly represented by the Editorial Board members, by the initiators of the journal, and, last but not least, the primary contribution of authors and reviewers. And readers who are the prospective authors.

---

[4] According to The Guardian, "In 2012 and 2013, Elsevier posted profit margins of more than 40%" (Buranyi 2017).
[5] https://www.coalition-s.org/
[6] Also called "Diamond Open Access" or "Gold Open Access without APC"

| What | Description | Resources |
| --- | --- | --- |
| Submission System: strategy | Decision-making process related to the detection of the right platform. The chosen submission system is based on multi-university initiative developing (free) open source software called Open Journal Systems[7] | EiC, DEiCs |
| Submission System: installation & maintenance | Operating environment preparation and software installation and configuration. Maintenance upon requests from the EiC | EiC Assistant (Dr. Juri Di Rocco) |
| Submission System: workflow configuration, user management | The workflow behind the submission, review, and production processes needs to be configured together with the notification email texts | EiC, DEiCs, EiC Assistant (Dr. Juri Di Rocco) |
| Website: renewal & extensions | Website face-lifting, new features and contents | EiC, DEiC |
| Website: maintenance | Content updates | EiC, DEiC |
| Workflow: review process | Manuscript review process management and reviewer assignment | EiC, DEiC, EB Members, Reviewers |
| Workflow: editorial assignment | Manuscript supervision assignment to members of the Editorial BOard | EiC, DEiC |
| Workflow: production | Camera-ready management, metadata (DOI) definition | EiC, DEiC |
| Workflow: publishing | Metadata registration (Crossref), content update | EiC, DEiC |
| Publicity: information material | Different formats have been produced and distributed, including flyers, stickers, banners, etc | EiC, DEiC |
| Publicity: social media | A twitter account is maintained and tweets are posted on a regular base advertising new initiatives and featured articles (most downloaded, most visited, etc) | EiC, DEiC |
| Publicity: youtube channel | A youtube channel has been created to collect video of paper presentation to increase visibility | EiC, DEiC |
| Publicity: certificates | Certification of appreciations for EB Members and Reviewers | EiC, DEiC |

**Table 1** Non-monetary cost distribution at JOT

### Acknowledgments

We are grateful to Loli Burgueño and Antonio Vallecillo for revising the initial version of the journal template and for the useful comments and suggestions that allowed us to considerably improve it.

### About the authors

**Alfonso Pierantonio** is professor at the Università degli Studi dell'Aquila (Italy) and Editor-in-Chief of the Journal of Object Technology. You can contact him at alfonso.pierantonio@univaq.it or visit http://pieranton.io.

**Mark van den Brand** is professor at the Technical University of Eindhoven (The Netherlands) and Deputy Editor-in-Chief of the Journal of Object Technology. You can contact him at m.g.j.v.d.brand@tue.nl or visit https://www.tue.nl/en/research/researchers/mark-van-den-brand/.

**Benoit Combemale** is professor at the University of Toulouse (France) and Deputy Editor-in-Chief of the Journal of Object Technology. You can contact him at benoit.combemale@irisa.fr or visit https://www.irit.fr/~Benoit.Combemale/.

### Acknowledgments

We are grateful to Loli Burgueño and Antonio Vallecillo for revising the initial version of the journal template and for the useful comments and suggestions that allowed us to considerably improve it.

### References

Bohannon, J. (2016). Who's downloading pirated papers? everyone. *Science (New York, NY)*, 352(6285):508.

Buranyi, S. (2017). Is the staggeringly profitable business of scientific publishing bad for science. *The Guardian*, 27(7):1–12.

Morais, R. and Borrell-Damian, L. (2018). Open Access 2016–2017 EUA Survey Results. *Retrieved August 2018*, 30:2018.

Van, N. (2013). R." open access: The true cost of science publishing. *Nature news feature*, pages 03–27.